\documentclass[conference]{IEEEtran}
\IEEEoverridecommandlockouts
\usepackage{cite}
\usepackage{amsmath,amssymb,amsfonts}
\usepackage{algpseudocode}
\usepackage{graphicx}
\usepackage{textcomp}
\usepackage{xcolor}
\usepackage{orcidlink}
\usepackage{hyperref}
\usepackage[capitalise]{cleveref}
\usepackage[para]{footmisc}
\usepackage{multirow}
\usepackage{ragged2e}
\usepackage{array}
\usepackage{pifont}
\usepackage{MnSymbol}
\usepackage{makecell}
\def\BibTeX{{\rm B\kern-.05em{\sc i\kern-.025em b}\kern-.08em
    T\kern-.1667em\lower.7ex\hbox{E}\kern-.125emX}}

\newcommand{\valid}{\ding{51}}
\newcommand{\invalid}{\ding{55}}
\newcommand{\inverse}{$\rightleftarrows$}

\usepackage{enumitem}
\usepackage{csquotes}

\makeatletter
\def\ps@IEEEtitlepagestyle{%
  \def\@oddfoot{\mycopyrightnotice}%
  \def\@oddhead{\hbox{}\@IEEEheaderstyle\leftmark\hfil\thepage}\relax
  \def\@evenhead{\@IEEEheaderstyle\thepage\hfil\leftmark\hbox{}}\relax
  \def\@evenfoot{}%
}
\def\mycopyrightnotice{%
  \begin{minipage}{\textwidth}
  \centering \scriptsize
  Copyright~\copyright~2023 IEEE. Personal use of this material is permitted. Permission from IEEE must be obtained for all other uses, in any current or future media, including\\reprinting/republishing this material for advertising or promotional purposes, creating new collective works, for resale or redistribution to servers or lists, or reuse of any copyrighted component of this work in other works by sending a request to pubs-permissions@ieee.org.
  \end{minipage}
}
\makeatother

\begin{document}

\title{Cloud-Native Architectural Characteristics and their Impacts on Software Quality: A Validation Survey}

\author{\IEEEauthorblockN{1\textsuperscript{st} Robin Lichtenthäler}
\IEEEauthorblockA{\textit{Distributed Systems Group} \\
	\textit{University of Bamberg, Germany}\\
	\orcidlink{0000-0002-9608-619X} 0000-0002-9608-619X
}
\and
\IEEEauthorblockN{2\textsuperscript{nd} Jonas Fritzsch}
\IEEEauthorblockA{\textit{Institute of Software Engineering} \\
    \textit{University of Stuttgart, Germany} \\
    \orcidlink{0000-0002-6121-2731} 0000-0002-6121-2731
}
\and
\IEEEauthorblockN{3\textsuperscript{rd} Guido Wirtz}
\IEEEauthorblockA{\textit{Distributed Systems Group} \\
	\textit{University of Bamberg, Germany}\\
	\orcidlink{0000-0002-0438-8482} 0000-0002-0438-8482
}
}

\maketitle

\begin{abstract}
Cloud-native architectures are often based on microservices and combine different aspects that aim to leverage the capabilities of cloud platforms for software development.
Cloud-native architectural characteristics like patterns and best practices aim to design, develop, deploy, and operate such systems efficiently with minimal time and effort. 
However, architects and developers are faced with the challenge of applying such characteristics in a targeted manner to improve selected quality attributes.
Hence, we aim to investigate relationships, or more specifically \textit{impacts}, between architectural characteristics of cloud-native applications, and quality aspects.
The architectural characteristics in consideration are based on our recently proposed quality model for cloud-native software architectures.
To validate its elements and revise this literature-based quality model, we conducted a questionnaire-based survey among 42 software professionals.
While the survey results reinforce the quality model to a fair extent, they also indicate parts requiring a revision.
Thus, as an additional contribution, we present an updated version of the quality model incorporating the survey results.
Practitioners will benefit from our work when designing and developing cloud-native applications in a quality-oriented way.
Researchers will moreover profit from our specifically developed questionnaire-based survey tool, which allows surveying complex structures like a hierarchical quality model.
\end{abstract}

\begin{IEEEkeywords}
cloud-native, survey, quality model validation
\end{IEEEkeywords}

\section{Introduction}

\noindent
Cloud-native describes an approach to building web applications which fully exploit the benefits of modern cloud environments.
It is seen as correlated with the topic of microservices, a popular architectural pattern for implementing service-oriented systems in cloud-native environments\cite{Baresi2019}.
While cloud-native as a term has been widely adopted in the industry and recently also in academic literature, existing definitions from industry \cite{CNCF2018} and academia \cite{Kratzke2017} still leave room for interpretation which complicates a commonly accepted understanding of the term. 
The main reason we see for this is that cloud-native covers a broad range of aspects from application design to deployment \cite{Lichtenthaeler2022}.
We thus see the need for a more distinct and structured definition of the concept of cloud-native applications to enable evaluations of software architectures and assess their conformance to cloud-native characteristics. 
To this end, our overarching goal is establishing a quality model for cloud-native software architectures.

In our previous work, we proposed an initial version of this quality model \cite{Lichtenthaeler2022}.
It structures architectural characteristics, also called \textit{product factors}, in a hierarchy of \textit{quality aspects}.
Within this hierarchy, \textit{impacts} describe how the presence of a product factor influences quality aspects.
This terminology is derived from the Quamoco Quality Meta Model \cite{Wagner2012} which serves as a conceptual foundation.
The initial quality model is based on existing studies and would also require additional tooling to be practically applicable for assessing software architectures. 
Before implementing such tooling, we consider it important to empirically validate the model with practitioners.
Such a validation would ensure that the described characteristics are relevant and understandable, and that their stated impacts on respective quality aspects correspond to phenomena observable in practice.
This is a key aspect of conceptual models, to which quality models belong, for them to be applicable in practice \cite{Moody2005}.
Validation can be seen as a form of formative evaluation \cite{Guzman2017} which is part of the creation process of a quality model or tool to ensure its soundness and reliability.
Several ways exist for performing validations, but not all are applicable in all phases of the creation process of a quality model \cite{Lichtenthaeler2022a}.
Especially in the earlier phases, when not all parts of a quality model are specified in detail, literature-based or empirical approaches using surveys or interviews are most suitable \cite{Lichtenthaeler2022a}.

Due to the breadth of the topic of cloud-native and the resulting complexity of the quality model, we investigate characteristics of cloud-native applications in a more general way, with the validation of the quality model being one intended purpose.
Hence, in this study, we conduct a questionnaire-based survey to validate the impacts stated in the initial quality model, as well as to identify additional impacts not previously considered.
We combine these goals in the following research question that guides our work: 

\begin{enumerate}[label=\textbf{RQ:},leftmargin=*]
    \item How do architectural characteristics of cloud-native software architectures impact different quality aspects?
\end{enumerate}

\noindent
To answer this question, we build on the architectural characteristics described in our previously proposed quality model \cite{Lichtenthaeler2022}.
An overview of this quality model is presented in section \ref{sec:model} where we also formulate more specific aims for its validation.
In section \ref{sec:relatedwork}, we review related work, such as other approaches for validating quality models, especially through surveys.
In section \ref{sec:methodology}, we describe our survey design and execution, and subsequently present its results in section \ref{sec:results}. 
Based on these results, we validate and revise our quality model in section \ref{sec:revised}.
Within our discussion in section \ref{sec:discussion} we outline potential future work and state threats to validity in section \ref{sec:threats}, before we conclude our work in section \ref{sec:conclusion}.

\section{Cloud-Native Quality Model}
\label{sec:model}

\begin{figure}[ht]
\centerline{\includegraphics{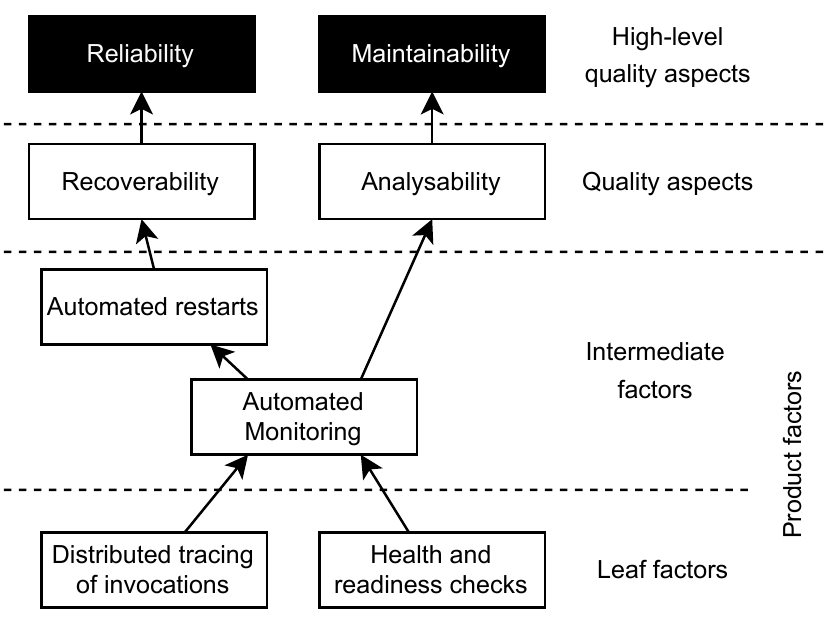}}
\caption{Small extract from the quality model with describing terms}
\label{fig:model-extract}
\end{figure}

\noindent
To build a foundation and common understanding for the approach used in this work, we present relevant aspects of the quality model in consideration.
\cref{fig:model-extract} shows a section of the quality model \cite{Lichtenthaeler2022} with a description of the elements on the right.
It is a hierarchical quality model \cite{Bansiya2002}, meaning that factors are ordered in a hierarchy of quality aspects which they impact.
\textit{High-level quality aspects} and \textit{quality aspects} are adopted from the ISO 25010 standard \cite{ISO/IEC2014} and \textit{product factors} were formulated during the creation of the initial quality model.
The arrows connecting the factors represent an \textit{impacts} relationship.

\noindent
The quality model covers multiple quality aspects, which largely match those of the ISO 25010 standard.
Factors can impact multiple quality aspects, with the general idea that also tradeoffs between quality aspects can be expressed. 
Below the quality aspects, there can be a hierarchy of factors which illustrate the structure of architectural characteristics and how they are interrelated.
In this way, we differentiate between \textit{intermediate factors} which impact and are impacted by other factors, and \textit{leaf factors} which are not impacted by further factors.
Leaf factors, thus, need to be measurable through metrics.
To enable uniform measurements, these metrics are defined based on a model for cloud-native software architectures, as investigated in \cite{Durr2022}. 
Intermediate factors can either also have metrics assigned to them or merely act as mediating and aggregating factors.
Hence, the overall model describes a graph.
However, edges can not be added arbitrarily.
To avoid ambiguities when applying the model, there should only be a single path from one leaf factor to a high-level quality aspect.

Based on the presented characteristics of the initially formulated quality model, we can now formulate the three goals for its validation which guide our validation approach and survey design:
\begin{enumerate}
    \item Confirming the impacts between factors and quality aspects
    \item Identifying additional impacts between factors and quality aspects which were not yet considered
    \item Determining the strength of impacts
\end{enumerate}

\section{Related Work}
\label{sec:relatedwork}

\noindent
Various approaches for creating quality models have been presented in literature, as reviewed by Nistala et al. \cite{Nistala2019}.
An important but not always explicitly considered step is the validation of a newly formulated quality model.
By validating a quality model, it is ensured that its theoretical concepts can be confirmed and explained also by a different type of rationale.
For example, if a quality model has been formulated based on literature, different types of rationale could be an empirical perspective from domain experts or quantitative data from benchmarking experiments.
Existing literature on performing validations of quality models has already been reviewed by Yan et al.\cite{Yan2017} who focus on commonly used model elements. 
In the same way, our previous work \cite{Lichtenthaeler2022a} discusses different types of rationale used for the formulation and validation of quality models.
For example, Mayr et al.~\cite{Mayr2012} developed a quality model for embedded systems and validated it by comparing the measurements gained through its application with quality ratings from independent experts. 
Such a validation, however, is not yet possible with the quality model considered in this work because this requires the quality model to be fully implemented.
Instead, we pursue an approach similar to the works of Khomh and Guéhéneuc~\cite{Khomh2008} or Gerpheide et al.~\cite{Gerpheide2015} who used questionnaires to measure impacts of software design patterns, respectively software properties on quality aspects. 

Compared to their works, ours differs in terms of the domain (cloud-native) and type of quality model (hierarchical).
Additionally, surveys based on questionnaires in the context of cloud computing or microservices have so far often been explorative, considering technologies, patterns, or approaches used by practitioners.
For example, Sousa et al. \cite{Sousa2021} investigated how cloud software engineering patterns are used in practice and Wang et al. \cite{Wang2021} investigated the relevance and usage of technologies and practices for microservices.
Our work is similar to these approaches, since we also seek to capture a perspective from practice on theoretical concepts in the context of cloud computing. 
But, in comparison, our goal is less explorative and more focused on the specific concepts of the quality model in consideration.

\section{Methodology}
\label{sec:methodology}

\noindent
Replicability is of particular importantance for survey research~\cite{CaterSteel2005}.
In this section, we therefore explain our process of survey design and execution.
We followed the survey research process as presented by Kasunic~\cite{Kasunic2005}, who lists seven stages for conducting a survey:

\begin{enumerate}
    \item Identify research objectives - see \ref{subsec:preparation}
    \item Identify \& characterize target audience - see \ref{subsec:preparation}
    \item Design sampling plan - see \ref{subsec:setup}
    \item Design \& write questionnaire - see \ref{subsec:setup}
    \item Pilot test questionnaire - see \ref{subsec:setup}
    \item Distribute the questionnaire - see \ref{subsec:distribution}
    \item Analyze results and write report - see \ref{subsec:distribution}
\end{enumerate}

\noindent
In addition, we describe in \ref{subsec:incorporation} how we used the survey results to continue our work on the quality model.

\subsection{Identification of Factors and Impacts to Validate}
\label{subsec:preparation}

\noindent
Our research objective is to validate and revise the elements of our quality model as described at the end of \cref{sec:model}.
The breadth of the topic resulted in a large number of factors for the quality model, namely 76.
This poses a challenge to our survey design, as the time and effort potential participants are willing to invest is usually limited.
We therefore adjust our objective by reducing the number of 76 original factors for the survey.
To do this in a systematic way, we:

\begin{enumerate}
    \item consider only the direct impacts from factors on quality aspects (e.g., \textit{Distributed tracing of invocations} $\rightarrow$ \textit{Analysability} in \cref{fig:model-extract}), excluding impacts from factors on mediating factors. (e.g., \textit{Distributed tracing of invocations} $\rightarrow$ \textit{Automated Monitoring} in \cref{fig:model-extract}
    \item exclude intermediate factors merely serving as \enquote{placeholders} (e.g. \textit{Automated Monitoring} in \cref{fig:model-extract}, but not \textit{Automated restarts})
    \item exclude factors which are less specific to cloud-native applications (e.g., \textit{Data encryption in transit})
\end{enumerate}

\noindent
This reduces the original number of 76 factors to 45, in addition to 24 quality aspects which can be impacted.

As our target audience, we identify IT professionals (e.g., developers, software engineers, software architects, IT managers) who have experience with implementing and operating web-based applications that run on cloud infrastructures. 
We regard their professional experience as the key enabler to reliably rate impacts of factors on quality aspects.
Still, we consider it a rather inclusive criterion, which we can adjust to avoid being too restrictive.

\subsection{Survey Setup and Pilot Study}
\label{subsec:setup}

\noindent
Our achievable sample size of the target audience is difficult to estimate in advance.
We therefore did not formulate an explicit sampling plan, but instead planned to distribute the survey via appropriate channels to achieve a maximum reach and collect as many answers as possible.
To ensure that participants belong to the target audience, an explicit description of the target audience was presented on the welcome page.

For the questionnaire design, we used a symmetric bipolar scale with a neutral alternative \cite{DeCastellarnau2017} for the impact ratings.
This allows us to measure the impact strength.
As too fine-grained scales have not been found to provide much benefit \cite{Lissitz1975}, we decided to use a 5-point scale with the following options for how a factor can impact a quality aspect:

\begin{itemize}
    \item factor has \textbf{positive impact (++)} on quality aspect  
    \item factor has \textbf{slightly positive impact (+)} on quality aspect
    \item factor has \textbf{no impact (0)} on quality aspect
    \item factor has \textbf{slightly negative impact (-)} on quality aspect
    \item factor has \textbf{negative impact (-{}-)} on quality aspect
\end{itemize}

\noindent
Using this scale would allow us to use an existing survey tool such as limesurvey\footnote{\url{https://www.limesurvey.org}} which we initially tried.
However, as we also wanted to have the possibility to detect previously not considered impacts, all potential impacts (i.e., 45 factors  * 24 quality aspects = 1080) would need to be rateable.
With the usual question layout from existing survey tools, this would make the survey overwhelming for participants.
Thus, we decided to build a custom survey frontend\footnote{\url{https://github.com/r0light/qmsurvey-frontend}} which aimed for a simple and intuitive rating by participants.
Each factor is presented in a uniform layout with a brief description.
In the same way, the potentially impacted quality aspects are presented below, grouped by their high-level quality aspects. 
By clicking on a quality aspect, an impact can be selected and rated. 
This design allows the participant to focus on a single factor at a time and choose all impacted quality aspects.
The consistent layout should help participants to quickly understand the process.

An additional difficulty was to estimate in advance the number of factors that participants would be willing to answer. 
In an ideal case, each participant would provide an answer to each factor (45), but the time that participants are willing to invest is usually very limited.
We thus tested our approach in a pilot study where participants could rate as many factors as they wanted. 
In this pilot, we asked them also to provide explicit feedback on the usability of the survey tool, the content, and the size of the survey.
The pilot study was conducted for two weeks in July 2022 with six participants who represented a small sample of the target audience.
We found that interested participants might be willing to provide more answers, while others might abort the survey soon if the number of questions is too high.
Hence, for the actual survey, we provided participants the possibility to rate as many factors as they would want to.
As well, we grouped factors thematically to make it easier for participants to choose them according to their topics of interest.
Finally, we included demographic questions about the job area, job title, industry sector, and years of experience with software development in general and experience with cloud computing in specific.

\subsection{Survey Distribution and Results Preparation}
\label{subsec:distribution}

\noindent
For the actual distribution of the survey, we sent the request for participation to:

\begin{itemize}
    \item professionals who had published articles addressing a related topic on popular blogging sites, such as Medium\footnote{\url{https://medium.com/}} 
    \item professionals who held a talk on a related topic at industry conferences, such as the CloudNativeCon\footnote{\url{https://www.cncf.io/kubecon-cloudnativecon-events/}}
    \item professionals who had published on a related topic at scientific conferences, such as the IEEE CLOUD\footnote{\url{https://conferences.computer.org/cloud}}
    \item professionals personally known to one of the authors
    \item community groups specifically considering cloud computing on social media sites, such as Reddit\footnote{\url{https://www.reddit.com/}}
\end{itemize}

\noindent
We preferred Email as the means of communication, but also relied on LinkedIn\footnote{\url{https://www.linkedin.com/}}, when contact details were not publicly accessible.
The survey was available online from October 2022 to December 2022.
During that timeframe, we received 42 complete submissions. 

We analyzed the results in two ways.
First, we used descriptive statistics to analyze how and for which factors participants provided answers.
Second, as our main analysis, we used the ratings for each combination of a factor and a quality aspect to compare them with impacts stated in the initial quality model.
As well, we were able to find additional impacts not covered by the original quality model.
To do so, and to also have an indicator for the significance of results, we used the \textit{Exact multinomial test of goodness-of-fit}\cite{McDonald2014} which is suitable when there are multiple values of one nominal variable (the different types of impact) and the sample size is small. 
In essence, the test compares the observed distribution of a variable with an assumed distribution representing the null hypothesis. 
The smaller the probability for an observed distribution under the null hypothesis is, the more significant the result is.
For the case of the null hypothesis, we assume that an impact can not be clearly stated, meaning that all ratings have the same probability. 
An exception to this is the \textit{no impact} rating which was selected as a default if no explicit rating was stated, and therefore we assume a twice as high probability for it.
That means, for the ratings \textbf{++}/\textbf{+}/\textbf{0}/\textbf{-}/\textbf{-{}-} we assume a ratio of \textbf{1}:\textbf{1}:\textbf{2}:\textbf{1}:\textbf{1} under the null hypothesis.
We summarized the analyses in a report which is available online\footnote{\url{https://github.com/r0light/qmsurvey-results}} and provides more details as well as the anonymized raw survey data.
This online report also includes demographic data on the background of participants which shows a mixed distribution of industry and academic backgrounds.

\subsection{Incorporation of the Results in the Quality Model}
\label{subsec:incorporation}

\noindent
With the survey report as a basis, we used those results that show significance to reiterate on the elements of the quality model.
That covers especially impacts which can not be validated through the survey results, so we reconsidered the literature initially used for formulating the quality model to restructure the hierarchy of factors where suitable. 
In addition, we iterated through the newly stated impacts and evaluated whether to include them in the quality model, again by reconsidering the initially used literature.
For factors with many impacts on several quality aspects, we reviewed whether they cover several distinctive aspects and can therefore be split up into separate factors.
And we reconsidered factors with unclear impacts, whether they can be defined and formulated more clearly.

This results in an updated and improved version of the quality model, presented in \cref{sec:revised}.
Finally, the results of the survey also highlight aspects to consider for further validation, for example where an impact could not be clearly derived.
We use these cases to formulate possibilities for future research.

\section{Survey Results}
\label{sec:results}

\subsection{General Descriptive Statistics}
\label{subsec:generalResults}

\begin{figure}[ht]
\centerline{
    \includegraphics[width=0.25\textwidth]{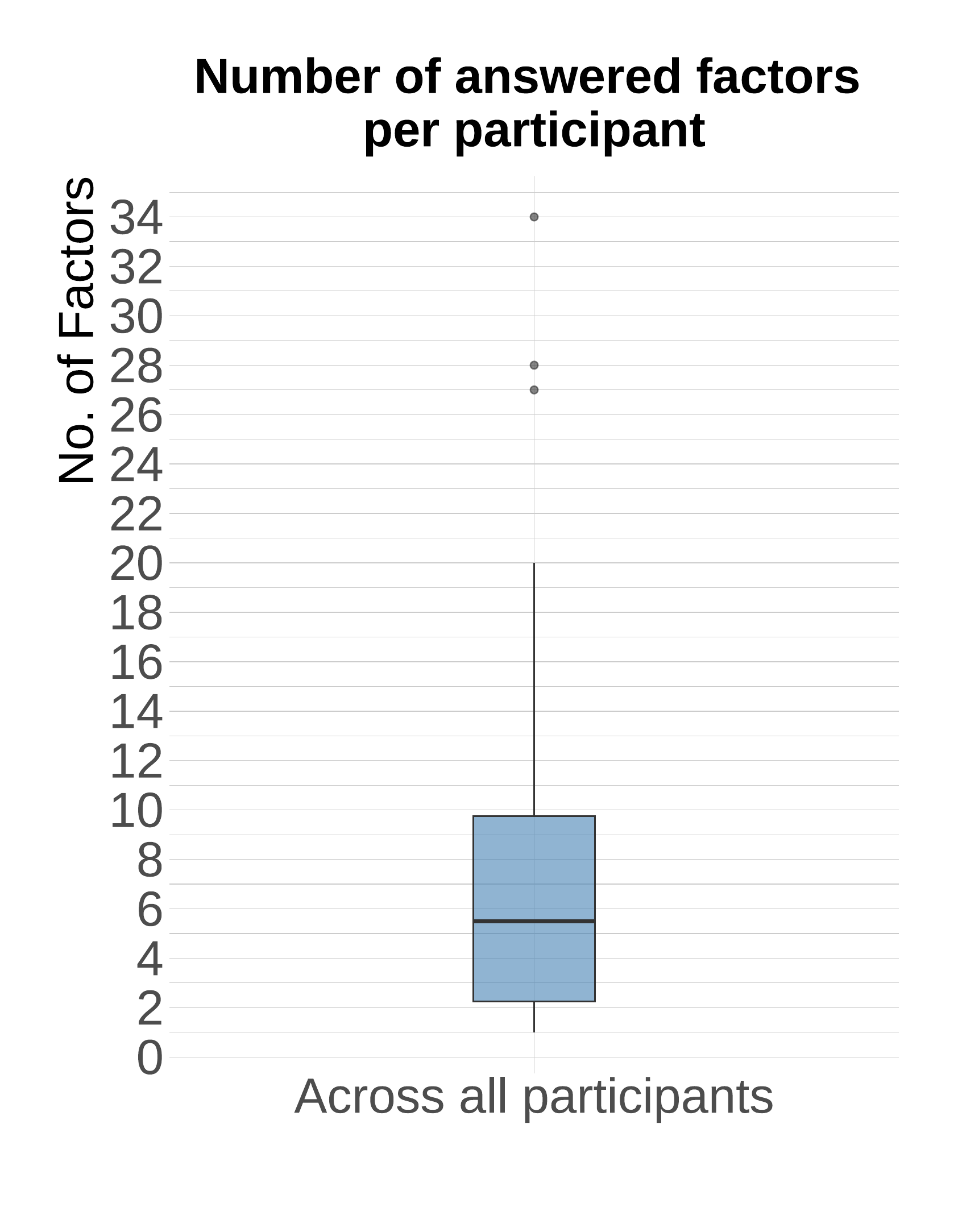}
    \includegraphics[width=0.25\textwidth]{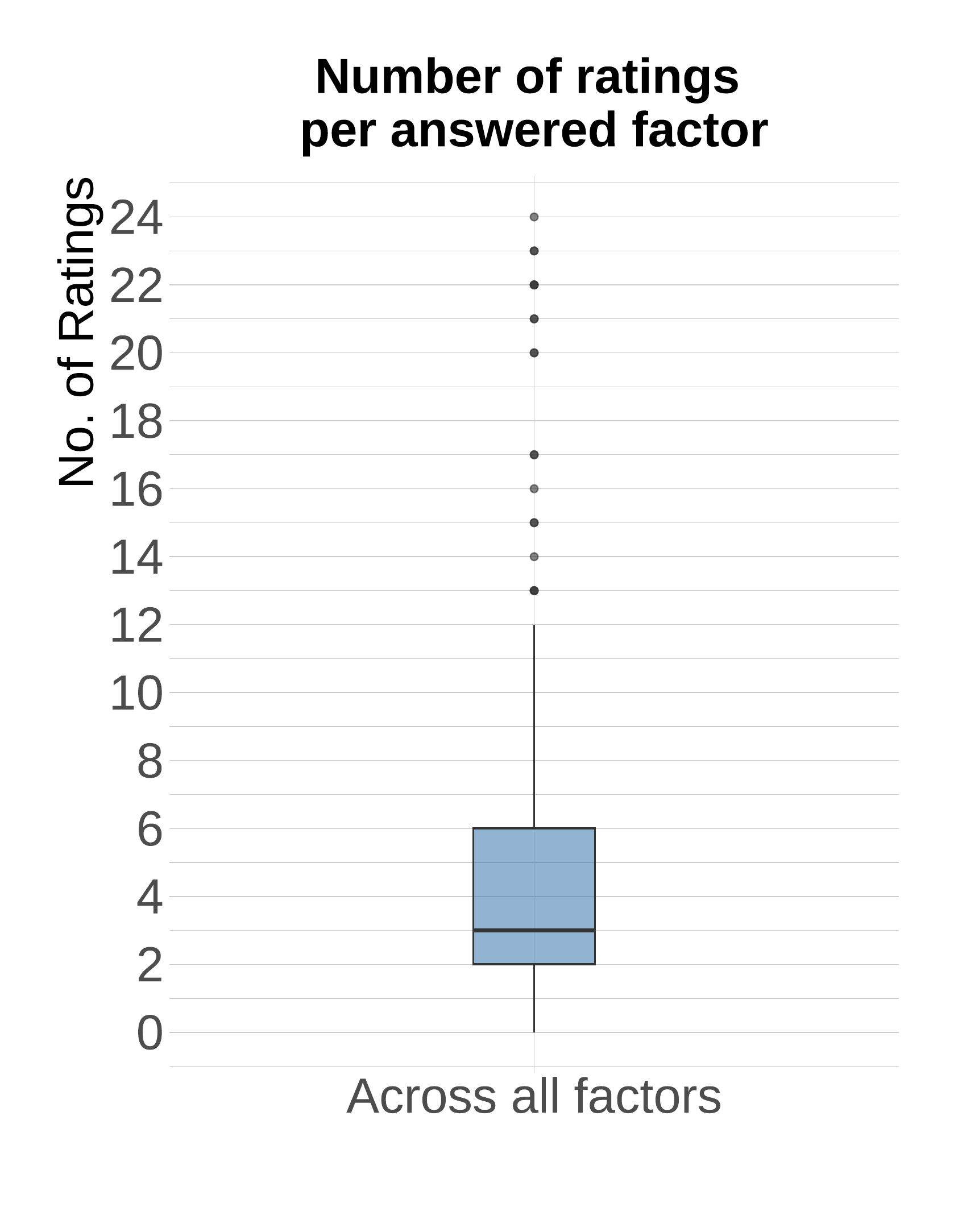}}
\caption{Statistical characteristics of the provided answers}
\label{fig:answerDistribution}
\end{figure}

\noindent
The flexible survey design allowed participants to freely choose how many factors they would rate.
We first counted for how many factors participants did provide answers and how many quality aspects they rated per factor (see \cref{fig:answerDistribution}).

\noindent
The median for the number of rated factors per participant is 5.5 which is fairly low compared to the number of 45 factors in total. 
Interestingly, there were also participants who provided answers to significantly more factors than this median, resulting in a wide variance of the overall number of answers per factor. 
While some factors like \textit{Automated infrastructure} or {Built-in autoscaling} were answered 20, respectively 17, times, others such as \textit{Specialized stateful services} or \textit{Separation by gateways} received only 2 answers. 
We need to take this variance into consideration for the interpretation of the results, since a higher number of answers provides a more reliable basis for interpretation.

For the number of quality aspects to rate per factor, the median is 3 which means that most participants stayed within the range we suggested in the survey (between 1 and 5).
But there are numerous outliers where participants stated many more impacts, meaning impacts other than \textit{no impact} for different quality aspects. 
In one case, a participant stated impacts for all 24 quality aspects, which might be the result of a misunderstanding of how to fill out the survey.

\subsection{Impact Validation}
\label{subsec:impactValidation}

\noindent
The variance in answered quality aspects per factor results in an overall high variation of stated impacts: Out of the 1080 potentially rateable impacts, 629 received at least one rating other than \textit{no impact}.
Thus, to evaluate which results are significant and therefore considered important for the quality model, we used a statistical test as described in \cref{subsec:distribution}.
Furthermore, we calculated for each factor-quality aspect combination the mean value out of the collected ratings by using [\textbf{+2},\textbf{+1},\textbf{0},\textbf{-1},\textbf{-2}] as the corresponding values for the rating options [\mbox{\textbf{++},\textbf{+},\textbf{0},\textbf{-},\textbf{-{}-}}].
This mean value together with the p-Value from the mentioned test build the basis for our interpretation of the results.
The \textit{mean} value expresses the type and strength of an impact, while the \textit{pValue} is an indicator for the significance of an impact stated by participants.
As the significance level we set $ \alpha=0.1 $ as a commonly chosen value.
For validating the initially stated impacts, we use the algorithm shown as a function in \cref{lst:validation} with \textit{impact} as a String to describe the initially stated impact being either \textit{``positive''} or \textit{``negative''}.
Furthermore, we set, as an additional significance indicator, a threshold of 5 for the total number of answers for a factor.
If the number of answers is above this threshold, \textit{aboveT} is \textit{true}, and \textit{false} otherwise.
The function returns a validation result in the form of a String which can be either \mbox{``\valid''} if the impact is valid, \mbox{``\invalid''} if the impact is invalid, or \mbox{``\inverse''} if the impact is found to be inverse (e.g., negative instead of positive).
The validation result is put in parentheses to mark it as uncertain, if either the p-Value is above the significance level or the number of answers for this factor is below the mentioned threshold.

\begin{figure}[ht]
\begin{algorithmic}[1]
\Procedure{validate}{$impact,mean,aboveT,pValue$}
\State $pThreshold \gets 0.1$ \Comment{The significance level}
\State $result \gets ``"$
\If{$mean < 0.5 \textbf{ and } mean > -0.5$}
  \If{$aboveT \textbf{ and } pValue < pThreshold$}
    \State $result \gets ``$\invalid$"$ 
  \Else \State $result \gets ``($\invalid$)"$ 
  \EndIf
\Else
  \If{$mean > 0$}
    \State $trend \gets ``positive"$
  \Else \State $trend \gets ``negative"$
  \EndIf
    \If{$trend = impact$}
      \State $result \gets ``$\valid$"$
    \Else
      \State $result \gets ``$\inverse$"$
    \EndIf
  \If{$\textbf{not}(aboveT \textbf{ and } pValue < pThreshold)$}
    \State $result \gets ``(" + result + ``)"$
  \EndIf
\EndIf
\State $\textbf{return } result$
\EndProcedure
\end{algorithmic}
\caption{Impact validation function}\label{lst:validation}
\end{figure}

\noindent
The result of applying this validation function is shown in \cref{tab:surveyresults}.
However, due to space constraints, not all factors are included, and can instead be found in the mentioned online repository.
The table also includes an indicator for the strength of an impact as either \textbf{-{}-}, \textbf{-}, \textbf{0}, \textbf{+}, or \textbf{+{}+}.
This strength indicator is based on the calculated mean value.
Furthermore, \cref{tab:surveyresults} includes impacts not considered in the initial quality model, but found to be significant for further consideration in the quality model.
These impacts thus have the value \textit{n/a} for the hypothesized type of impact and its validation.

\begin{table*}
    \centering
    \caption{Survey results: Factor-Quality aspect combinations with p $<$ 0.1}
    \label{tab:surveyresults}
\begin{tabular}[ht]{>{\raggedright\arraybackslash}m{14.5em}|>{\raggedright\arraybackslash}m{8.5em}|l|l|c|c|c|c|c|c|c|c}
\hline
\textbf{Factor} & \textbf{Quality Aspect} & \textbf{hypothesized} & \textbf{-{}-} & \textbf{-} & \textbf{0} & \textbf{+} & \textbf{++} & \textbf{Mean} & \textbf{Impact} & \textbf{Validation} & \textbf{p-Value}\\
\hline
Addressing abstraction & Modularity & positive & 0 & 0 & 6 & 0 & 0 & 0.00 & 0 & \invalid & 0.0696\\
\cline{1-12}
Authentication delegation & Authenticity & positive & 0 & 0 & 0 & 2 & 3 & 1.60 & ++ & \valid & 0.0288\\
\cline{1-12}
 & Modifiability & positive & 0 & 2 & 12 & 1 & 5 & 0.45 & + & \invalid & 0.0224\\
\cline{2-12}
\multirow{-2}{14.5em}{\raggedright\arraybackslash Automated infrastructure} & Recoverability & positive & 0 & 0 & 7 & 3 & 10 & 1.15 & + & \valid & 0.0006\\
\cline{1-12}
Automated restarts & Recoverability & positive & 0 & 0 & 2 & 2 & 8 & 1.50 & ++ & \valid & 0.0007\\
\cline{1-12}
Backing service decentralization & Co-existence & positive & 0 & 0 & 6 & 0 & 0 & 0.00 & 0 & \invalid & 0.0696\\
\cline{1-12}
 & Elasticity & positive & 0 & 0 & 3 & 2 & 12 & 1.53 & ++ & \valid & 0.0000\\
\cline{2-12}
\multirow{-2}{14.5em}{\raggedright\arraybackslash Built-in autoscaling} & Resource utilization & positive & 0 & 0 & 3 & 3 & 11 & 1.47 & + & \valid & 0.0000\\
\cline{1-12}
Cloud vendor abstraction & Adaptability & positive & 0 & 0 & 4 & 1 & 6 & 1.18 & + & \valid & 0.0176\\
\cline{1-12}
Consistent centralized metrics & Analyzability & positive & 0 & 0 & 1 & 3 & 3 & 1.29 & + & \valid & 0.0560\\
\cline{1-12}
 & Analyzability & positive & 0 & 0 & 1 & 2 & 5 & 1.50 & ++ & \valid & 0.0122\\
\cline{2-12}
\multirow{-2}{14.5em}{\raggedright\arraybackslash Distributed tracing of invocations} & Recoverability & positive & 0 & 0 & 7 & 0 & 1 & 0.25 & + & \invalid & 0.0423\\
\cline{1-12}
 & Modifiability & positive & 0 & 0 & 6 & 0 & 0 & 0.00 & 0 & \invalid & 0.0696\\
\cline{2-12}
\multirow{-2}{14.5em}{\raggedright\arraybackslash Dynamic scheduling} & Recoverability & positive & 0 & 0 & 6 & 0 & 0 & 0.00 & 0 & \invalid & 0.0696\\
\cline{1-12}
Health and readiness Checks & Recoverability & positive & 0 & 0 & 5 & 0 & 4 & 0.89 & + & \valid & 0.0414\\
\cline{1-12}
Horizontal data replication & Time-behaviour & positive & 0 & 0 & 7 & 0 & 1 & 0.25 & + & \invalid & 0.0423\\
\cline{1-12}
Infrastructure abstraction & Adaptability & positive & 0 & 0 & 2 & 4 & 2 & 1.00 & + & \valid & 0.0970\\
\cline{1-12}
Limited request trace scope & Modifiability & positive & 0 & 0 & 6 & 0 & 0 & 0.00 & 0 & \invalid & 0.0696\\
\cline{1-12}
Managed backing services & Resource utilization & positive & 0 & 0 & 7 & 2 & 0 & 0.22 & + & \invalid & 0.0433\\
\cline{1-12}
 & Simplicity & positive & 0 & 0 & 11 & 1 & 2 & 0.36 & + & \invalid & 0.0093\\
\cline{2-12}
\multirow{-2}{14.5em}{\raggedright\arraybackslash Managed infrastructure} & Resource utilization & positive & 0 & 0 & 7 & 4 & 3 & 0.71 & + & \valid & 0.0765\\
\cline{1-12}
Physical service distribution & Availability & positive & 0 & 0 & 1 & 1 & 8 & 1.70 & ++ & \valid & 0.0001\\
\cline{1-12}
Rolling upgrades enabled & Availability & positive & 0 & 0 & 0 & 0 & 3 & 2.00 & ++ & (\valid) & 0.0185\\
\cline{1-12}
Service replication & Time-behaviour & positive & 0 & 0 & 4 & 4 & 6 & 1.14 & + & \valid & 0.0140\\
\cline{1-12}
 & Modifiability & positive & 0 & 0 & 12 & 3 & 1 & 0.31 & + & \invalid & 0.0048\\
\cline{2-12}
 & Installability & positive & 0 & 0 & 8 & 1 & 7 & 0.94 & + & \valid & 0.0039\\
\cline{2-12}
\multirow{-3}{14.5em}{\raggedright\arraybackslash Use infrastructure as code} & Recoverability & positive & 0 & 0 & 8 & 3 & 5 & 0.81 & + & \valid & 0.0378\\
\cline{1-12}
Vertical data replication & Time-behaviour & positive & 0 & 0 & 1 & 0 & 4 & 1.60 & ++ & \valid & 0.0288\\
\Xhline{1.5\arrayrulewidth}
\cline{1-12}
 & Modularity & n/a & 0 & 1 & 2 & 0 & 7 & 1.30 & + & n/a & 0.0022\\
\cline{2-12}
 & Reusability & n/a & 0 & 0 & 4 & 1 & 5 & 1.10 & + & n/a & 0.0480\\
\cline{2-12}
\multirow{-3}{14.5em}{\raggedright\arraybackslash API-based communication} & Replaceability & n/a & 0 & 0 & 7 & 0 & 3 & 0.60 & + & n/a & 0.0377\\
\cline{1-12}
 & Modularity & n/a & 0 & 0 & 13 & 2 & 5 & 0.60 & + & n/a & 0.0031\\
\cline{2-12}
 & Reusability & n/a & 0 & 0 & 11 & 2 & 7 & 0.80 & + & n/a & 0.0028\\
\cline{2-12}
 & Testability & n/a & 0 & 2 & 9 & 3 & 6 & 0.65 & + & n/a & 0.0997\\
\cline{2-12}
 & Capability & n/a & 0 & 0 & 14 & 1 & 5 & 0.55 & + & n/a & 0.0008\\
\cline{2-12}
 & Elasticity & n/a & 0 & 0 & 11 & 2 & 7 & 0.80 & + & n/a & 0.0028\\
\cline{2-12}
 & Resource utilization & n/a & 0 & 0 & 13 & 2 & 5 & 0.60 & + & n/a & 0.0031\\
\cline{2-12}
 & Time-behaviour & n/a & 0 & 0 & 13 & 2 & 5 & 0.60 & + & n/a & 0.0031\\
\cline{2-12}
 & Installability & n/a & 0 & 1 & 8 & 2 & 9 & 0.95 & + & n/a & 0.0050\\
\cline{2-12}
 & Availability & n/a & 0 & 0 & 7 & 3 & 10 & 1.15 & + & n/a & 0.0006\\
\cline{2-12}
\multirow{-10}{14.5em}{\raggedright\arraybackslash Automated infrastructure} & Fault tolerance & n/a & 0 & 0 & 11 & 1 & 8 & 0.85 & + & n/a & 0.0006\\
\cline{1-12}
 & Availability & n/a & 0 & 1 & 0 & 3 & 8 & 1.50 & ++ & n/a & 0.0001\\
\cline{2-12}
\multirow{-2}{14.5em}{\raggedright\arraybackslash Automated restarts} & Fault tolerance & n/a & 0 & 0 & 3 & 3 & 6 & 1.25 & + & n/a & 0.0137\\
\cline{1-12}
 & Capability & n/a & 0 & 0 & 7 & 3 & 7 & 1.00 & + & n/a & 0.0095\\
\cline{2-12}
 & Time-behaviour & n/a & 0 & 0 & 8 & 4 & 5 & 0.82 & + & n/a & 0.0292\\
\cline{2-12}
 & Availability & n/a & 0 & 0 & 8 & 5 & 4 & 0.76 & + & n/a & 0.0292\\
\cline{2-12}
\multirow{-4}{14.5em}{\raggedright\arraybackslash Built-in autoscaling} & Fault tolerance & n/a & 0 & 0 & 8 & 5 & 4 & 0.76 & + & n/a & 0.0292\\
\cline{1-12}
 & Reusability & n/a & 0 & 0 & 6 & 1 & 4 & 0.82 & + & n/a & 0.0804\\
\cline{2-12}
\multirow{-2}{14.5em}{\raggedright\arraybackslash Cloud vendor abstraction} & Replaceability & n/a & 0 & 0 & 6 & 1 & 4 & 0.82 & + & n/a & 0.0804\\
\cline{1-12}
Command Query Responsibility Segregation & Simplicity & n/a & 0 & 4 & 3 & 0 & 0 & -0.57 & - & n/a & 0.0560\\
\cline{1-12}
Dynamic scheduling & Capability & n/a & 0 & 0 & 1 & 0 & 5 & 1.67 & ++ & n/a & 0.0027\\
\cline{1-12}
Health and readiness Checks & Availability & n/a & 0 & 0 & 3 & 1 & 5 & 1.22 & + & n/a & 0.0414\\
\cline{1-12}
 & Elasticity & n/a & 0 & 0 & 7 & 4 & 3 & 0.71 & + & n/a & 0.0765\\
\cline{2-12}
 & Availability & n/a & 0 & 0 & 4 & 3 & 7 & 1.21 & + & n/a & 0.0074\\
\cline{2-12}
 & Fault tolerance & n/a & 0 & 0 & 9 & 2 & 3 & 0.57 & + & n/a & 0.0499\\
\cline{2-12}
 & Maturity & n/a & 0 & 0 & 8 & 3 & 3 & 0.64 & + & n/a & 0.0765\\
\cline{2-12}
\multirow{-5}{14.5em}{\raggedright\arraybackslash Managed infrastructure} & Recoverability & n/a & 0 & 0 & 8 & 3 & 3 & 0.64 & + & n/a & 0.0765\\
\cline{1-12}
Mostly stateless services & Testability & n/a & 0 & 0 & 2 & 0 & 7 & 1.56 & ++ & n/a & 0.0006\\
\cline{1-12}
Physical data distribution & Fault tolerance & n/a & 0 & 0 & 1 & 1 & 5 & 1.57 & ++ & n/a & 0.0095\\
\cline{1-12}
Physical service distribution & Fault tolerance & n/a & 0 & 0 & 2 & 3 & 5 & 1.30 & + & n/a & 0.0233\\
\cline{1-12}
 & Availability & n/a & 0 & 0 & 1 & 4 & 9 & 1.57 & ++ & n/a & 0.0000\\
\cline{2-12}
\multirow{-2}{14.5em}{\raggedright\arraybackslash Service replication} & Fault tolerance & n/a & 0 & 0 & 5 & 3 & 6 & 1.07 & + & n/a & 0.0238\\
\cline{1-12}
Usage of existing solutions for non-core capabilities & Simplicity & n/a & 0 & 0 & 3 & 4 & 0 & 0.57 & + & n/a & 0.0560\\
\cline{1-12}
 & Reusability & n/a & 0 & 0 & 9 & 2 & 5 & 0.75 & + & n/a & 0.0242\\
\cline{2-12}
 & Testability & n/a & 0 & 0 & 9 & 3 & 4 & 0.69 & + & n/a & 0.0404\\
\cline{2-12}
 & Adaptability & n/a & 0 & 0 & 8 & 6 & 2 & 0.62 & + & n/a & 0.0163\\
\cline{2-12}
 \multirow{-4}{14.5em}{\raggedright\arraybackslash Use infrastructure as code} & Availability & n/a & 0 & 0 & 9 & 3 & 4 & 0.69 & + & n/a & 0.0404\\
\hline
\end{tabular}
\end{table*}

As the table shows, several initially stated impacts could be validated.
And for many, there is at least a tendency towards their validation recognizable. 
By tendency, we mean that individual ratings supported these impacts, but there are simply not enough answers to rate the impact as validated based on our evaluation.
Although not shown in \cref{tab:surveyresults}, this data is also available in the mentioned online repository.

There are also several impacts which could not be validated and hence might have to be removed from the quality model. 
This can be partly explained by the fact that we did not consider mediating factors in the survey which have an explanatory function.
When they are missing, the indirect impact from a factor on a quality aspect might be less obvious.
For example, the impact from \textit{Limited request trace scope} on \textit{Modifiability} is mediated by \textit{Service independence} which was not part of the survey. 
Therefore, our results need to be considered on a case-by-case basis in regard to deciding whether to remove or keep the impact in the quality model.
Nevertheless, removing invalidated impacts could also make the quality model less complex and therefore more understandable.

Reconsidering the extract from the quality model in \cref{fig:model-extract}, we can see how the survey results help to clarify impact relationships.
For \textit{Health and readiness Checks}, a positive impact on \textit{Recoverability} was stated by the participants, but not on \textit{Analysability}.
While for \textit{Distributed Tracing of invocations}, a positive impact on \textit{Analysability} was stated, the hypothesized positive impact on \textit{Recoverability} could not be validated. 

\begin{figure}[ht]
\centerline{
    \includegraphics[width=0.25\textwidth]{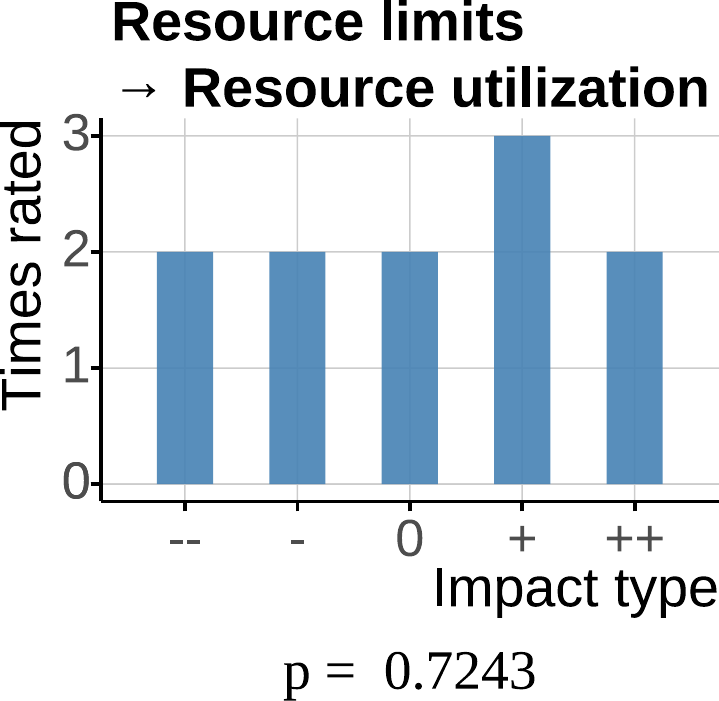}
    \includegraphics[width=0.25\textwidth]{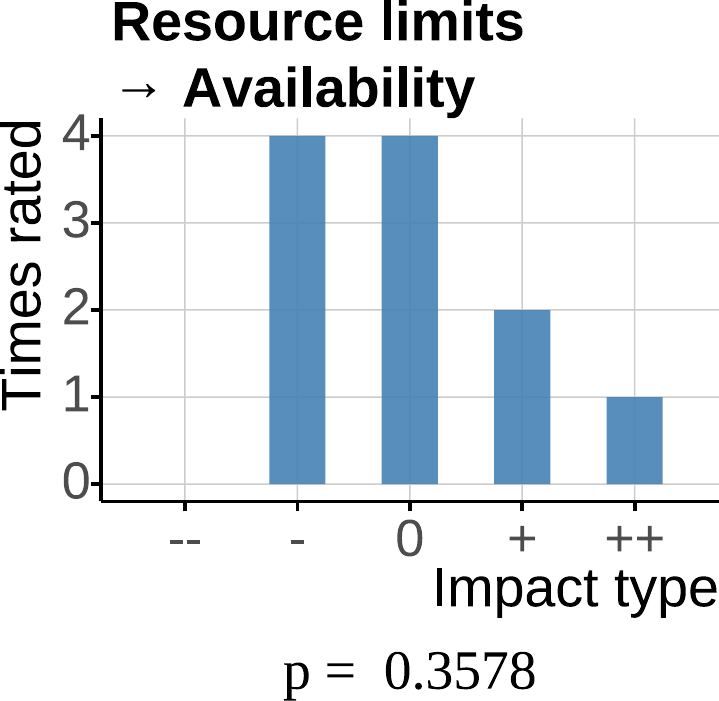}}
\caption{Ambiguous results for the factor \textit{Resource limits}}
\label{fig:resourceLimits}
\end{figure}

\noindent
Furthermore, there are a few results which show a mixed picture of stated impacts.
These are interesting in that they reveal factors which can be understood differently and therefore need to be revised.
As it can be seen in \cref{fig:resourceLimits}, this is the case for the factor \textit{Resource limits} which we described in the survey as: \enquote{\textit{For all components the maximum amount of resources a component can consume is limited based on its predicted needs so that resources are provisioned efficiently. That means a component gets the resources that it needs, but not more than necessary. By making the resource requirements explicit, for example in a configuration file, these limits can be enforced by the infrastructure.}}
We assume that we had made the implicit assumption here that resource requirements of components are known well so that appropriate resource limits can be set.
If appropriate resource limits are set, resource utilization is impacted positively because components can be provided with the resources that they need, and the infrastructure only has to provide the specified resources, not more.
However, if that assumption does not hold and resource limits are set too strict, a negative impact on resource utilization can result.
Components might not get the resources they actually need, even though resources would still be available which are then in turn also not utilized.
A similar situation is possible regarding the impact on availability.
When appropriate resource limits are set, misbehaving components can not take away shared resources from other components.
This protects these other components and therefore has a positive impact on availability.
But the availability of individual components might be impacted negatively, if they do not get the resources they need to function properly due to too strict resource limits.
Thus, the quality model needs to be revised at this point to better reflect these circumstances.

These results cover the first aim of validating impacts in the initially stated quality model, as described in \cref{sec:model}. 

\subsection{Analysis of Additional Impacts}
\label{subsec:additionalImpacts}

\noindent
The above discussed impact of \textit{Resource limits} on \textit{Availability} is one of the impacts which we had not included in the initial quality model.
Identifying such additional impacts was the second aim of our survey.
The large variance of provided answers represents a challenge for this aim too. 
Hence, it is necessary to make a selection of impacts which should be considered for incorporation in the quality model.
For this selection, we again rely on the significance indicators of the gained results.
In \cref{tab:surveyresults}, the additional impacts with the highest significance, as indicated by the p-Value, are included.
But there are numerous additional impacts for which the results are slightly above the significance level, which might still be considerable, too.

What is also notable in these results is that there are factors such as \textit{Automated infrastructure}, \textit{Managed infrastructure}, or \textit{Use infrastructure as code} for which impacts on several quality aspects are found to be significant. 
This is even true for several quality aspects with the same high-level quality aspect.
For example, \textit{Automated infrastructure} has slightly positive impacts on \textit{Recoverability}, \textit{Availability}, and \textit{Fault tolerance}, which are all sub-aspects of \textit{Reliability}.
We see two possible explanations for this result.
One is that these factors are rather abstract and comprise several aspects, which makes their assessment ambiguous.
Depending on which aspects of a factor participants think of, their ratings may involve multiple quality aspects.
This could be addressed by splitting up these factors into (sub-)factors that capture more clearly distinguishable aspects of cloud-native applications and make the formulated impact relationships more precise.
Another possible explanation is that the quality aspects themselves are not completely conceptually disjoint.
Following this explanation, certain quality aspects overlap conceptually, i.e. they share certain concepts.
If a factor in question now impacts specifically such a shared concept, it would be plausible for it to impact both quality aspects simultaneously.
The consequence of this explanation would be that the different quality aspects could not only be considered as distinguishable sub-aspects of their respective high-level aspects, but rather as top-level quality aspects themselves.
In either case, these results of factors with impacts on many quality aspects highlight elements of the quality model that need to be revised in order to formulate a more conceptually valid quality model.

We find that the survey uncovered additional, previously not considered, impacts. 
For incorporating them into the quality model, a decision has to be made again on a case-by-case-basis.
The indicators presented in \cref{tab:surveyresults} serve as a basis in this regard.

To summarize, the results of our survey successfully contribute to the first two goals in \cref{sec:model}.
The third goal, evaluating the strength of impacts, is more difficult to support with the results.
Although we differentiate between strengths of impacts in the data, this differentiation is rather weak, simply because of the small overall number of answers per factor.
We therefore see no solid basis for comparing different impacts according to their strength, as well as for the incorporation of the results into the quality model.
Thus, in the following, we also do not differentiate further between \textit{positive} and \textit{slightly positive} or \textit{negative} and \textit{slightly negative} impact types.

\begin{figure*}[!ht]
\centerline{
    \includegraphics[width=0.87\textwidth]{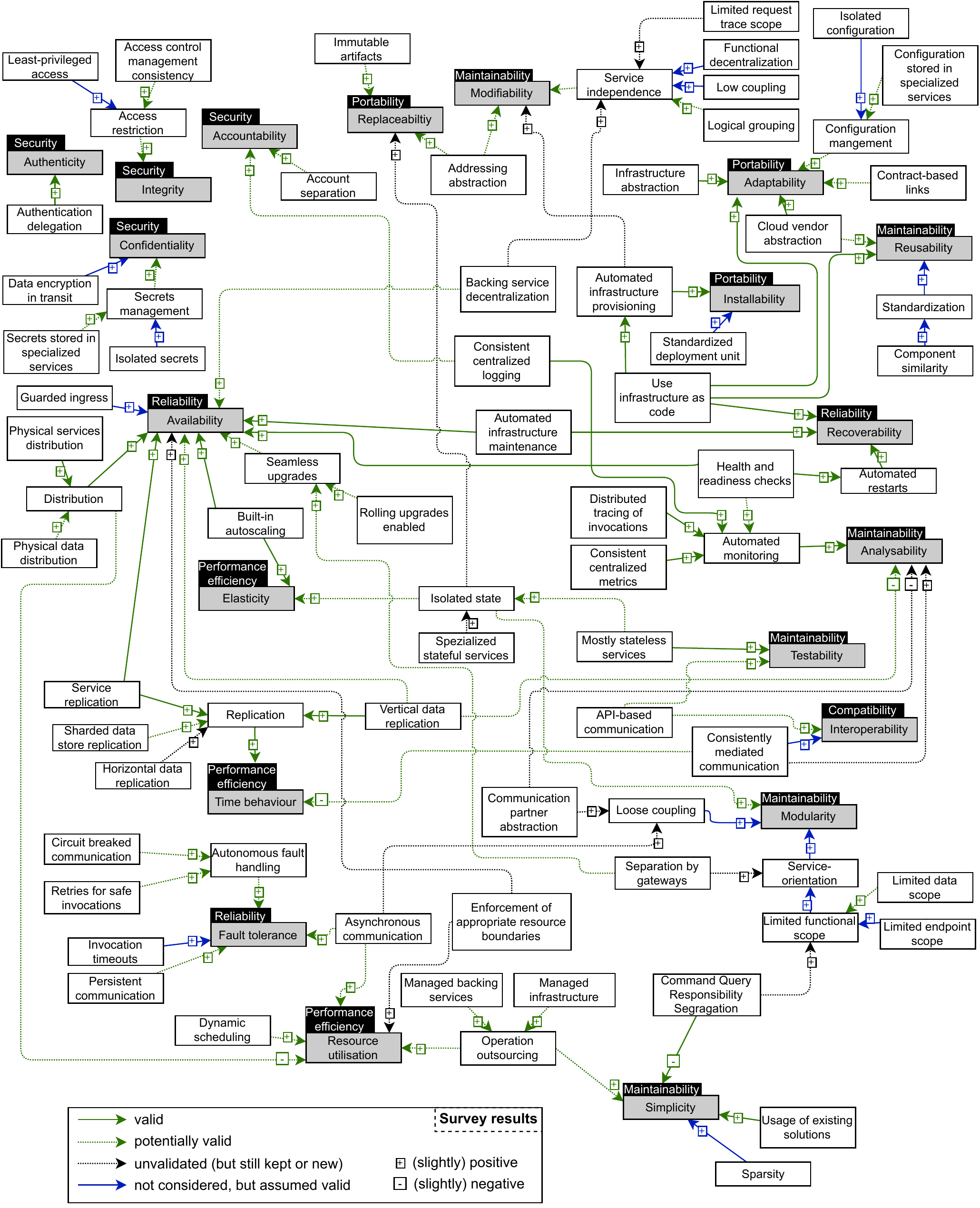}}
\caption{Revised Quality Model}
\label{fig:revisedQualityModel}
\end{figure*}

\section{Revised Quality Model}
\label{sec:revised}

\noindent
After the analysis of the survey results, we revised the quality model to incorporate the gained results and improve its conceptual soundness.
As mentioned in the introduction, this covers the quality aspects, product factors, and the impacts which connect them.
Metrics and evaluations of the quality model are not in the scope of this paper and hence excluded.

We started our revision with the initially stated impacts. 
Those which could be validated (``\ding{51}'') are kept in the quality model.
Those which could \textit{not} be validated (``\invalid'') are removed from the quality model.
Exceptions are cases where the investigated impact was mediated by a factor not considered in the survey.
Here, we decided to keep the impact for further validation if we still considered it relevant.
This applies for example to the impact from \textit{Limited request trace scope} on \textit{Modifiability}, mediated by the factor \textit{Service independence}.
For all impacts that show unclear results (``(\valid)'', ``(\invalid)'', or ``($\rightleftarrows$)''), we thoroughly reconsidered the impact to decide whether to keep it or remove it from the quality model.
To this end, we also reconsidered literature which represents the initial reason for that impact.
Furthermore, we revised factors that show mixed results by reformulating their descriptions or by separating them into more distinct (sub-)factors.
For example, we redefined the above-mentioned factor \textit{Resource limits} into a new factor \textit{Enforcement of appropriate resource boundaries} that captures both aspects of providing components with the resources they need, but also enforcing limits.

We continued our revision with the additional impacts identified through the survey.
Ordered by the significance from highest to lowest, we decided for each impact whether it can be incorporated into the quality model and, if mediating factors are involved, how exactly this can be done.
Additionally, when considering one impact from a factor, we also compared it to the other impacts stated for that factor to decide if also the factor itself needs to be revised.
This explicitly also includes those impacts with low significance because they potentially indicate a need to revise a factor.
For example, this is the case for the factor \textit{Automated infrastructure}, as already shown in \cref{subsec:additionalImpacts}.
In that case, we refined the factor \textit{Automated infrastructure} into the factors \textit{Automated infrastructure provisioning} and \textit{Automated infrastructure maintenance}.
\textit{Automated infrastructure provisioning} covers the aspect of providing requested infrastructure in an automated way, therefore having a positive impact on \textit{Installability}.
In contrast, \textit{Automated infrastructure maintenance} considers the automated management of infrastructure after it has been provisioned to ensure, for example, \textit{Availability} through automated updates.

During this revision process, we also had to acknowledge that factors do not always impact only one quality aspect of the same high-level quality aspect, as we had assumed in the initial quality model.
We therefore also decided to use the quality aspects directly as the top-level elements in the quality model, instead of the high-level quality aspects.
This way, factors can have impacts, even of different types, on different quality aspects associated with the same high-level quality aspect.
One example is the factor \textit{Command Query Responsibility Segregation}, for which we state a positive impact on \textit{Limited functional scope} and therefore \textit{Modularity}, but for which the survey also shows a negative impact on \textit{Simplicity}.
Another example is \textit{Use infrastructure as code} which positively impacts \textit{Portability} through \textit{Installability} as well as \textit{Adaptability}.

The outcome of incorporating the survey results is the revised quality model shown in \cref{fig:revisedQualityModel}, also available online\footnote{\url{https://r0light.github.io/cna-quality-model/}}.
The results of the survey are reflected by marking the impacts based on their significance.
In conclusion, this revised quality model is now based on both literature and empirical evidence through our survey.
As such, the quality model's conceptual basis has been strengthened compared to its initial version.

\section{Discussion}
\label{sec:discussion}

\noindent
In this section, we discuss some more general insights gained throughout the process of conducting the survey and provide an answer to our initially stated research question.
When preparing and conducting the survey, we faced two major challenges.
First, the comparatively large scope of the quality model and second, the aim of identifying impacts between factors and quality aspects not previously considered.
To address them, we developed a customized survey tool.
While the tool as such proved to be effective, its flexibility led to a large variance in how the survey was filled out by participants, as shown in \cref{subsec:generalResults}.
Some participants answered only a few factors, others answered more than expected, which is helpful and would not have been possible with a fixed number of factors to answer.
Regarding the number of quality aspects rated per factor, however, the large variance resulting from the flexibility made it difficult to interpret the results.
We consider this a general challenge for empirically validating large models.
Restricting the number of impact ratings per factor, or asking specifically about the impacts stated in the initial quality model, would have been viable alternatives too. 
They, however, bear the risk of introducing unnecessary bias.
To avoid such bias, we wanted to explicitly not restrict the number of possible impact ratings per factor.
As well, we did not treat the impacts stated in the initial quality model differently than other potential impacts.

A constraint of the questionnaire we used is that participants stated impact ratings, without having to provide a rationale for their ratings.
Especially for factors with many stated impacts or unclear impact types, this would have been helpful as an explanation.
But having to fill out an additional text field per rating would have also meant an additional effort, which we wanted to avoid.
For this, interviews would be more suitable to investigate the rationale behind impacts in-depth, as it has been done for example by Lampasona et al.~\cite{Lampasona2013}.
Such interviews could be employed in potential future work.
Especially those relationships between factors and quality aspects with mixed or less clear results could be picked up in interviews.

As a more general outcome, we can state that the survey results do not provide a definitive validation of the quality model, but leave room for further work.
One direction of future work, we aim to do, is using experiments to investigate impacts based on metrics.
For that, suitable metrics need to be selected that are able to measure the presence of a factor in a software architecture at design time.
If such metrics can be correlated with metrics that measure corresponding quality aspects at runtime, this also validates impacts from factors on quality aspects.
Metrics, that measure factors corresponding to architectural characteristics, are called \textit{architectural metrics}.
To define and establish such architectural metrics, a suitable representation of software architectures in the context of the quality model is required.
We have already investigated suitable modeling approaches for software architectures of cloud-native applications \cite{Durr2022}.
Based on our there presented modeling approach, we plan to make the mentioned metrics measurable and therefore conduct such experiments.

Reconsidering our initially formulated research question on how architectural characteristics of cloud-native applications impact quality aspects, we can state that our survey results show some obvious impacts, for example from \textit{Automated restarts} on \textit{Recoverability} or from \textit{Physical service distribution} on \textit{Availability}.
But there are also many less clear results where, e.g., impacts from one factor on many quality aspects have been found.
This highlights the factors that need to be revised and potentially split up into separate factors having distinct impacts on certain quality aspects.
But it also indicates that the quality aspects from the ISO 25010 standard \cite{ISO/IEC2014} are not always conceptually disjoint, i.e. free from overlaps.
From a strict perspective, this hinders their usage in hierarchical quality models.
This is because, one factor should not have impacts on the same top-level element via different intermediate elements (i.e., via several paths), as stated by AL-Badareem et al.~\cite{ALBadareen2015} who formulated rules for the quality of hierarchical quality models.
If the high-level aspects, such as \textit{Maintainability} are used as top-level elements, for \textit{API-based communication} impacts on both \textit{Modularity} and \textit{Reusability} were found, but both are sub-aspects of \textit{Maintainability}.
To address this problem in the revised quality model, we decided to use the different quality aspects as the top-level element, instead of the high-level quality aspects.

In conclusion, the survey enabled us to improve the quality model through adding and removing impacts as well as the refinement of factors. 
It is thus an important supplement to the overall formulation process of the quality model.

\section{Threats to Validity}
\label{sec:threats}

\noindent
Several limitations to our work need to be considered.
The participants were self-recruited, meaning that although we distributed the survey in a controlled way, it was open to anyone interested, in contrast to using personalized invitations.
To mitigate the risk that participants would be insufficiently qualified, we described the intended target group at the beginning of the survey so that each participant was made aware of it.
While the overall number of 42 participants was rather low, we are still convinced to have attracted highly qualified professionals due to the mentioned precautions. 
Still, we need to state the limited external validity.
The queried participants can not be considered representative for the whole target group, and our results are thus not generalizable.
To ensure construct validity, that means whether the factors are understandable and relevant, we used a pilot study to ask for feedback on the individual factors and reformulated them when necessary \cite{Kasunic2005}.
Nevertheless, due to the novelty of the topic of cloud-native, we could not rely on validated factor descriptions from other work.
Thus, factor descriptions may still contain ambiguities and could have been understood in different ways.

\section{Conclusion}
\label{sec:conclusion}

\noindent
In this work, we conducted a questionnaire-based validation survey to investigate relationships between architectural characteristics of cloud-native applications and quality aspects, using a hierarchical quality model.
In summary, our work includes a three-fold contribution:
1) We present our approach for applying a questionnaire-based survey to empirically validate a hierarchical quality model.
This includes the design of a custom survey tool for the flexible rating of impacts between factors and quality aspects, guidelines for conducting such a survey, and a method for the analysis of the received submissions.
2) We present the results from the survey, which reflect clear impacts between certain architectural characteristics and quality aspects, but also characteristics that are to some extent ambiguous. 
3) We present an updated version of our quality model which incorporates the results from our survey.
It explains the relationships between architectural characteristics and quality aspects in a structured way.
The thematic breadth of cloud-native is covered by the consideration of multiple quality aspects and by the formulation of distinctive factors to capture architectural characteristics.
Significant parts of the quality model have now been validated and improved. 
This work thereby establishes a solid foundation for our future activities in completing this quality model for the evaluation of software architectures according to cloud-native characteristics.
Also because of the mentioned limitations, in future work we plan to further improve the applicability of the model and validate its elements from additional perspectives.

\bibliography{qmsurvey}
\bibliographystyle{IEEEtran}

\end{document}